\documentclass[twocolumn,prb,showpacs,preprintnumbers,amsmath,amssymb]{revtex4}

\usepackage{natbib}

\usepackage{graphicx}
\usepackage{dcolumn}
\usepackage{bm}
\usepackage[T1]{fontenc}
\usepackage{multirow}
\usepackage{color}

\begin{document}

\title{Superconductivity and Electron Correlations in Kagome Metal LuOs$_3$B$_
2$}

\author{Yusen Xiao$^{1,2}$, Qingchen Duan$^{2}$, Tao jia$^{1}$, Yajing Cui$^{1}$, Shaohua Liu$^{3}$, Zhiwei Wen$ ^{4}$, Liangwen Ji$^{5}$, Ruidan Zhong$^{2,3}$\footnote{E-mail:rzhong@sjtu.edu.cn}, Yongliang Chen$^{1}$\footnote{E-mail:llchen@home.swjtu.edu.cn }, Yong Zhao$^{1,6,7}$\footnote{E-mail:zhaoyong@fjnu.edu.cn } }

\address
{  
\\$^{1}$    Key Laboratory of Magnetic Levitation Technologies and Maglev Trains (Ministry of Education), School of Physical Science and Technology, Southwest Jiaotong University, Chengdu 610031, China  
\\$^{2}$ Tsung-Dao Lee Institute, Shanghai Jiao Tong University, Shanghai 201210, China  
\\$^{3}$ School of Physics and Astronomy, Shanghai Jiao Tong University, Shanghai, 200240, China  
\\$^{4}$ School of intelligent manufacturing, Sichuan University of Arts and Scienc, Dazhou 635000, china
\\$^{5}$ School of Physics, Zhejiang University, Hangzhou 310058, China 
\\$^{6}$ Fujian Provincial Collaborative Innovation Center for Advanced High-Field Superconducting Materials and Engineering, Fuzhou 350117, China  \\$^{7}$ College of Physics and Energy, Fujian Normal University, Fuzhou, Fujian 350117,China }

\date{\today}

\begin{abstract}

We report a comprehensive investigation of the physical properties of LuOs$_3$B$_2$, characterized by an ideal Os-based kagome lattice. Resistivity and magnetization measurements confirm the emergence of type-II bulk superconductivity with a critical temperature $T_\text{c} \approx 4.63$ K.  The specific heat jump and the calculated electron-phonon coupling parameter support a moderately coupled superconducting state. Electron correlation effects are supported  by the enhanced Wilson ratios. First-principles calculations reveal hallmark features of kagome band structure, including Dirac points, van Hove singularities, and quasi-flat bands,  primarily derived from the Os $d$ orbitals. The inclusion of spin-orbit coupling opens a gap at the Dirac points, significantly altering the electronic properties. Furthermore, the superconductivity and electronic properties of isomorphic compounds are discussed. This work provides a thorough exploration of the superconducting and normal states of LuOs$_3$B$_2$, deepening the understanding of kagome superconductors.

\end{abstract}

\pacs{XXX}

\maketitle
\section{\label{sec:level1}Introduction}

As a unique lattice structure characterized by two-dimensional corner-sharing triangles, the kagome lattice has emerged as a forefront platform in condensed matter physics owing to its capacity to stabilize exotic electronic states in quantum materials\cite{1,2,3}. This lattice's distinctive symmetry leads to the emergence of Dirac points, van Hove singularities (vHSs), and flat bands, which are pivotal in driving both topological and correlated electronic behaviors\cite{4,5,6}. Recent experimental progress has predominantly focused on transition metal-based kagome systems, such as Fe$_3$Sn$_2$\cite{Fe3Sn2,Fe3Sn2-2,Fe3Sn2-3}, FeSn\cite{FeSn,FeSn-2}, FeGe\cite{FeGe,FeGe2,FeGe3}, Co$_3$Sn$_2$S$_2$\cite{CSS,CSS2}, and  $R$$T$$_6$$X$$_6$($R$ = rare earth, $T$= V, Cr, Mn, Nb, $X$= Sn, Ge)\cite{GdNb,cr166, Mn166, Sc166, ThVSN}, revealing rich topological properties. However, the exploration of intertwined superconductivity (SC), electronic correlations, and band topology in kagome metals remains constrained by the lack of suitable material platforms.

A significant breakthrough was achieved with the discovery of superconducting and charge density wave (CDW) states in $A$V$_3$Sb$_5$ ($A$ = K, Rb, Cs)\cite{135-1,135-2,135-3}. These compounds exhibit additional intriguing phenomena, including paired density waves and anomalous Hall effects\cite{135PDW,135-hall}. Notably, the recently synthesized CsCr$_3$Sb$_5$ demonstrates moderate electronic correlations and pressure-induced unconventional superconductivity at approximately 4 GPa\cite{Cr135,Cr135-2,Cr135-3}. Another family of kagome superconductors, $R$$T$$_3$$X$$_2$ ($R$ = La, Y, Lu, Th; $T$ = Ru, Rh, Ir, Os; $X$ = Si, B, Ga), have also attracted considerable interest\cite{LaRuSi-1,LaRuFeSi,LaRuSiprb,LaRuSiwhh,LaRhB,LaIrGa,ThRuSi,YRSi}. Among these, LaRu$_3$Si$_2$ ($T_c \approx 7.8$ K) and YRu$_3$Si$_2$ ($T_c \approx 3.0$ K) have been extensively studied due to their prototypical kagome-derived band structure and strong electron correlations, likely arising from the kagome flat band\cite{LaRuSiwhh,YRSi}. However, $R$Ru$_3$Si$_2$ compounds exhibit a distorted kagome lattice, LuOs$_3$B$_2$ in the family stands out as an ideal structure of the kagome geometry\cite{ku1980superconducting}. This structural distinction positions LuOs$_3$B$_2$ as a potentially pristine platform for investigating kagome-related features in band structures and their interplay with superconductivity.

In this work, we report a systematic study of the superconducting and normal-state characteristics of LuOs$_3$B$_2$. Bulk superconductivity with a critical temperature $T_c = 4.63$ K is unambiguously established through resistivity, magnetic susceptibility, and specific heat measurements. Experimental results support a moderately coupled superconducting state consistent with the predictions of BCS theory. Interestingly, the enhanced Wilson ratio implies the existence of electron correlation effects in the normal state. First-principles calculations reveal further complexity in the electronic structure of LuOs$_3$B$_2$. The calculated band structure hosts multiple nontrivial topological crossings near the Fermi level, with Dirac points developing pronounced gaps  upon inclusion of spin-orbit coupling (SOC). These gap-opening mechanisms, coupled with the symmetry-protected band degeneracies along the $\Gamma$-M-K high-symmetry line, hint at the coexistence of potential nontrivial topological states and correlated electron physics in LuOs$_3$B$_2$.

\section{\label{sec:level2}Experimental}

A polycrystalline sample of LuOs$_3$B$_2$ was synthesized from high-purity powders of Lu (99.5\%), Os (99.5\%, Alfa), and B (99.99\%, Alfa). The powders were mixed in stoichiometric ratios using an agate mortar and cold-pressed into a pellet under an argon atmosphere within a glove box to ensure a controlled environment. The pellet was then arc-melted on a water-cooled copper hearth, undergoing four melting cycles with intermediate flipping to achieve compositional homogeneity, yielding a sample with a metallic luster.

X-ray diffraction (XRD) analysis was conducted using a Bruker D8 AdvanceEco diffractometer, with data collected over a $2\theta$ range of 10°-90°. The XRD patterns were refined using the Rietveld method implemented in the GSASII software. Electrical resistivity and magnetoresistance measurements were performed using a Quantum Design Physical Property Measurement System (PPMS) equipped with a standard four-probe configuration. DC magnetization measurements were carried out down to 1.8 K using a vibrating sample magnetometer (VSM) integrated with the PPMS. Specific heat capacity measurements were also conducted down to 1.8 K using the relaxation method within the PPMS.

First-principles calculations were performed using the Vienna ab initio simulation package (VASP) based on density functional theory (DFT)\cite{vasp}. The generalized gradient approximation (GGA) with the Perdew-Burke-Ernzerhof (PBE) exchange-correlation functional was employed\cite{GGA}. The projector augmented wave (PAW) pseudopotential method was used to enhance computational accuracy. Structural optimization was performed with a plane-wave cutoff energy of 500 eV and a $10 \times 10 \times 16$ Monkhorst-Pack $k$-point grid centered at the $\Gamma$ point to sample the Brillouin zone. Experimental lattice parameters were used as initial inputs, and the total energy was converged to within $1 \times 10^{-8}$ eV at each step. For the density of states (DOS) calculations, a denser $20 \times 20 \times 32$ Monkhorst-Pack $k$-point grid was utilized to ensure high precision in the electronic structure characterization.

\section{\label{sec:level3}Results and Discussion}

\begin{figure}[tbp]
\includegraphics[width=9 cm]{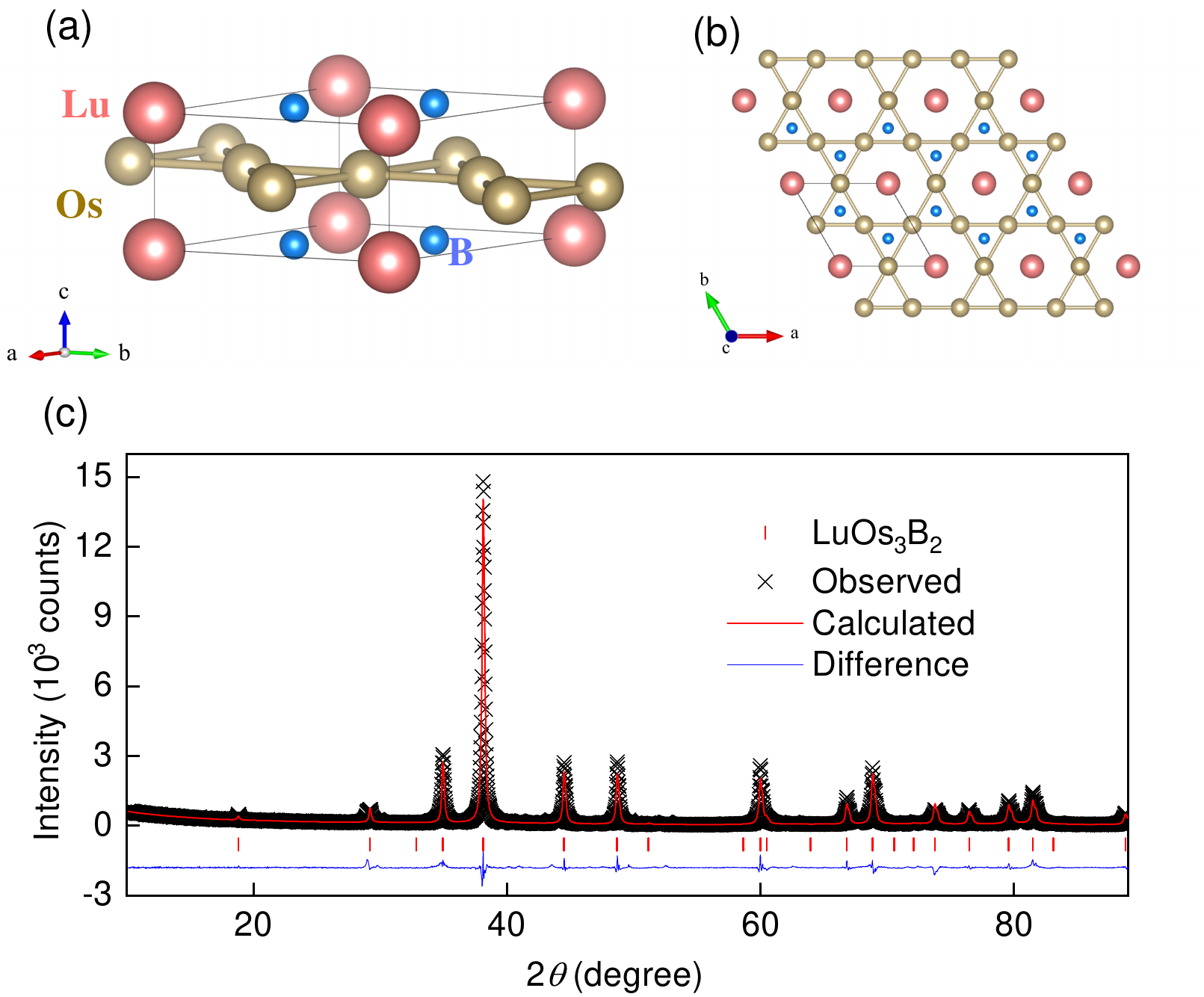}
\caption {(Color online) (a,b) Crystal structure of LuOs$_3$B$_2$, a side view, and a top view. (c) Rietveld refinement profile of LuOs$_3$B$_2$ polycrystalline sample.}
\label{fig1}
\end{figure}

\begin{table}[!h]
\caption{Crystallographic data of LuOs$_3$B$_2$ at room temperature.} \label{tab1}
\setlength{\tabcolsep}{0.8mm}
\renewcommand\arraystretch{1.2}
\begin{center}
\begin{tabular}{ccccccc}
\hline
\hline
&Compounds       &&& LuOs$_3$B$_2$ &   \\
&space group     &&& $P6/mmm$          & \\
&$a$ (\AA)     &&& 5.449(3)      & \\
&$c$ (\AA)     &&& 3.058(8)       & \\
&$V$ (\AA$^3$) &&& 78.67       & \\
&Os-Os distance (\AA) &&& 2.7246        & \\
&$R_{\rm{wp}}$ (\%)   &&& 9.31           & \\
&$\chi^2$             &&& 1.99           & \\
&$Z$             &&& 1               & \\
\hline
atom  &  site    &  $x$    &  $y$  &  $z$   & Occ. \\
\hline
Lu   &  1$a$      &  0    &   0   &   0      &  1&          \\
Os    &  3$g$      &  0.5   &  0  & 0.5    &  1   &          \\
B    &  2$c$      &  0.33333  & 0.66667  & 0   &  1     &       \\
\hline
\hline
\end{tabular}
\end{center}
\end{table}

 Figure \ref{fig1}(a) presents the XRD pattern for LuOs$_3$B$_2$, along with the Rietveld refinement profile indicative of the CeCo$_3$B$_2$-type structure, which belongs to the space group $P6/mmm$. The refined structural parameters are summarized in Table I. The lattice parameters are $a$ = 5.449(3) Å and $c$ = 3.058(8) Å, aligning with prior reports\cite{ku1980superconducting}. Figure \ref{fig1}(b) and (c) show the crystal structure of LuOs$_3$B$_2$, which consists of alternating Lu–B and Os layers stacked along the $c$-axis. Interestingly, the Os ions form perfect kagome layers along the crystallographic $c$-axis, in contrast to the distorted kagome layers observed in the isostructural compound LaRu$_3$Si$_2$.  The Os-Os distance within these kagome layers measures 2.7246 Å, comparable to V-V bond in the $A$V$_3$Sb$_5$ family\cite{ortiz2019new}. Note that the short $c$-axis suggests that interlayer coupling between the kagome planes could be enhanced.

Figure 2 shows the resistivity data of the LuOs$_3$B$_2$ sample. At 300 K, the zero-field resistivity is $126\,\mathrm{\mu\Omega\cdot cm}$, a value comparable to that of LaRh$_3$B$_2$\cite{LaRhB}. Upon cooling, the resistivity displays metallic behavior without anomalies. The residual resistivity ratio (RRR), defined as $\rho(300\,\mathrm{K})/\rho(5\,\mathrm{K})$, is approximately 3.5, indicating significant contributions from grain boundaries to resistivity variations. At high temperatures, the resistivity curve $\rho(T)$ exhibits a distinct sublinear behavior. This observation aligns with a recently proposed semiclassical theory, where such behavior is attributed to the interplay between Dirac cones and van Hove singularities near the Fermi surface\cite{RTsublinear,RTsublinear2}. Notably, similar phenomena have been reported in other kagome metals\cite{LaRuSiwhh,YRSi,RTsublinear2}. In the low-temperature range ($5\,\mathrm{K} < T < 40\,\mathrm{K}$), the $\rho(T)$ curve is well-described by the equation $\rho(T) = \rho_0 + AT^2 + BT^5$, where $\rho_0$ represents the residual resistivity, while the terms $AT^2$ and $BT^5$ arise from electron--electron and electron--phonon scattering, respectively. The fitting yields $\rho_0 = 36.1\,\mathrm{\mu\Omega\ cm}$, $A = 6.6 \times 10^{-3}\,\mathrm{\mu\Omega\ cm\ K^{-2}}$, and $B = 2.8 \times 10^{-8}\,\mathrm{\mu\Omega\ cm\ K^{-5}}$. Since $AT^2$ dominates over $BT^5$ below $50\,\mathrm{K}$, electron--electron scattering is the primary mechanism governing resistivity at low temperatures.

\begin{figure}[tbp]
	\includegraphics[width=8cm]{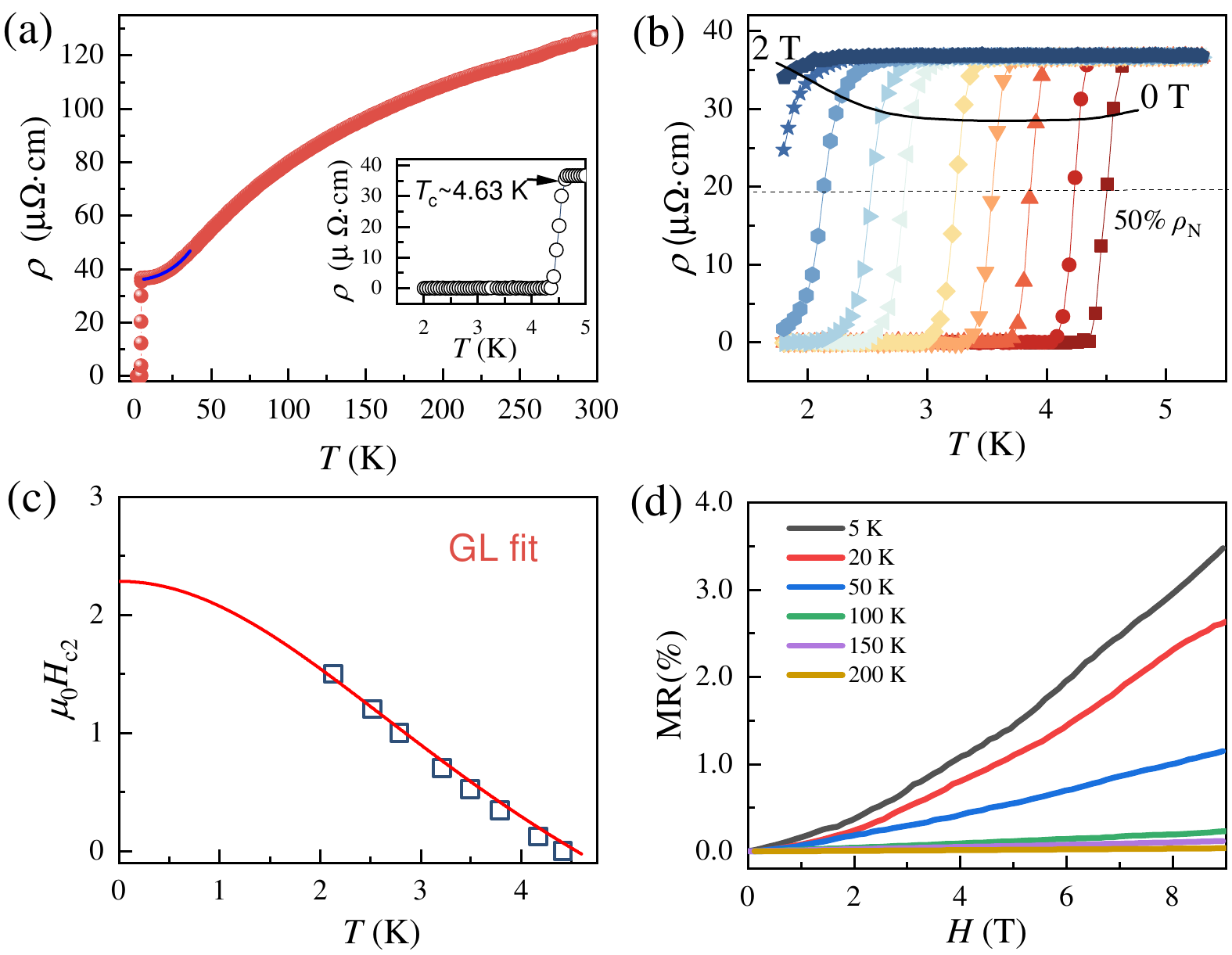}
	\caption {(Color online) (a) Temperature-dependent resistivity of LuOs$_3$B$_2$, with the inset illustrating the criteria for determining the onset of the resistivity drop. (b) Resistivity transition between 1.8 K and 5 K under applied magnetic fields ranging from 0 to 2 T. The horizontal dashed line at 50\% of the normal-state resistivity defines the superconducting transition temperature under field. (c) Upper critical field $\mu_0 H_{c2}(T)$ as a function of temperature, fitted using the Ginzburg-Landau theory (red line). (d) Magnetoresistance measured at selected temperatures.}
	\label{fig2}
\end{figure}

\begin{figure}[tbp]
\includegraphics[width=8.5cm]{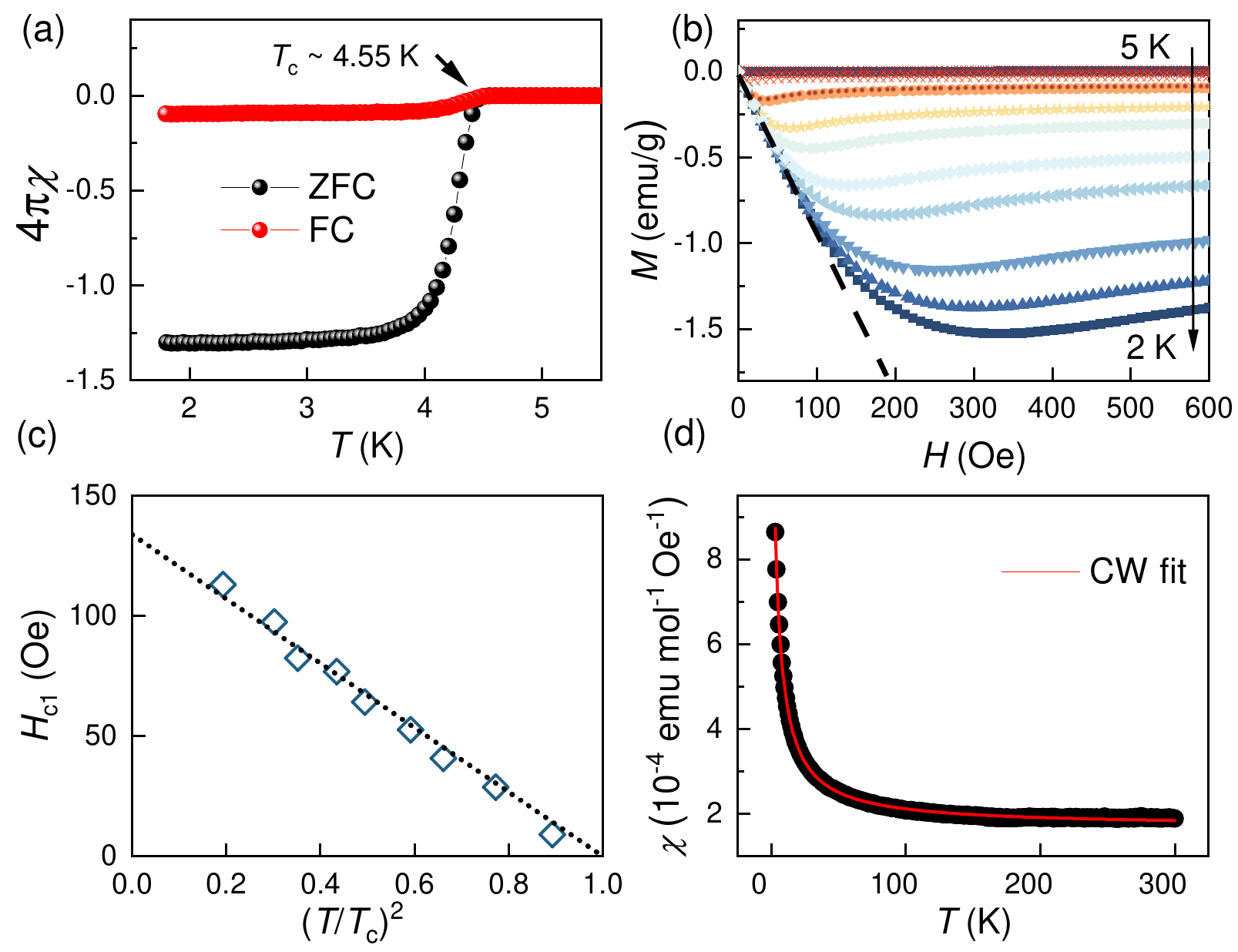}
\caption {(Color online) (a) Low-temperature susceptibility of LuOs$_3$B$_2$ in zero-field-cooled (ZFC, black) and field-cooled (FC, red) modes under a 10 Oe magnetic field. (b) Field-dependent initial magnetization at temperatures below $T_c$. (c) Temperature dependence of the lower critical field $H_{c1}(T)$. (d) Temperature dependence of magnetic susceptibility $\chi(T)$ at 1 T, with the red line indicating a Curie-Weiss fit.  }
\label{fig3}
\end{figure}

A superconducting transition is observed at $4.63\,\mathrm{K}$, marked by a sharp resistivity drop, consistent with previous reports. Figure 2(b) depicts the temperature dependence of resistivity under static magnetic fields ($0$ to $2\,\mathrm{T}$). Here, $T_c(H)$ is defined as the temperature at which resistivity drops to $50\%$ of the normal-state value. Using the Ginzburg--Landau (GL) formula $\mu_0 H_{c2}(T) = \mu_0 H_{c2}(0) \frac{T_{c,0}^2 - T^2}{T_{c,0}^2 + T^2}$, where $\mu_0 H_{c2}(0)$ is the zero-temperature upper critical field and $T_{c,0}$ is the superconducting transition temperature at zero field, the fit yields $\mu_0 H_{c2}(0) = 2.3\,\mathrm{T}$. This value is significantly below the Pauli limit $\mu_0 H_{c2}^P = 1.84 T_c = 6.95\,\mathrm{T}$, suggesting that orbital pair breaking is the dominant mechanism in LuOs$_3$B$_2$. The coherence length  can be calculated as $\xi$ = 120 Å using the relation $\mu_0 H_{c2}(0) = \frac{\Phi_0}{2\pi \xi^2}$, where $\Phi_0$ is the magnetic flux quantum. Additionally, magnetoresistance measurements up to $9\,\mathrm{T}$ at selected temperatures reveal a small, unsaturated positive MR, as shown in Fig. 2(d).

Figure \ref{fig3}(a) displays the temperature dependence of the magnetic susceptibility $\chi(T)$ at $H = 10\,\mathrm{Oe}$, measured using both zero-field-cooling (ZFC) and field-cooling (FC) protocols. A pronounced diamagnetic signal below $4.55\,\mathrm{K}$ confirms the superconducting transition observed in resistivity measurements. The superconducting volume fraction reaches $130\%$ at $1.8\,\mathrm{K}$ in zero-field-cooled (ZFC) mode, which is attributed to the demagnetization effect, indicating bulk superconductivity. Compared to the ZFC curve, the FC curve exhibits a weaker diamagnetic signal due to flux pinning, suggesting that LuOs$_3$B$_2$ is a type-II superconductor.  

Figure \ref{fig3}(b) shows the field dependence of the zero-field-cooled magnetization at various temperatures. The lower critical field $H_{c1}$ is determined from the point where the magnetization $M$ deviates from linearity. The temperature dependence of $H_{c1}$ is fitted using the Ginzburg-Landau (GL) formula $H_{c1}(T) = H_{c1}(0)\left[1 - \left(T/T_c\right)^2\right]$, where $H_{c1}(0)$, the lower critical field at $0\,\mathrm{K}$, is approximately $13.4\,\mathrm{mT}$, as shown in Fig. \ref{fig3}(c). From the relations $\mu_0 H_{c1}(0) = \frac{\ln(\lambda/\xi) \Phi_0}{4\pi \lambda^2}$ and $\kappa = \lambda/\xi$, the penetration depth and Ginzburg-Landau parameter are derived as $\lambda = 1830\,\mathrm{\AA}$ and $\kappa = 15.2$, respectively. The thermodynamic critical field at 0 K, $\mu_0 H_c(0)$, is estimated using the relation $H_{c1}(0) \cdot H_{c2}(0) = H_c^2(0) \ln \kappa$. For LuOs$_3$B$_2$, this yields $\mu_0 H_c(0) = 106 \, \text{mT}$. 

The magnetic susceptibility under a field of $1\,\mathrm{T}$ remains nearly temperature-independent above $200\,\mathrm{K}$, indicating a Pauli paramagnetic state. By fitting the data to the Curie--Weiss law $\chi(T) = \chi_0 + \frac{C}{T - \theta_W}$, we obtain $\chi_0 = 1.71 \times 10^{-4}$ emu mol$^{-1}$, $\theta_W = -2.21\,\mathrm{K}$, and $C = 0.004007$ emu K mol$^{-1}$. Using the relation $C = \frac{\mu_0 \mu_{\mathrm{eff}}^2}{3k_B}$, the Curie--Weiss constant $C$ corresponds to a small effective moment $\mu_{\mathrm{eff}} = 0.17\,\mu_\mathrm{B}/\mathrm{f.u.}$. The temperature-independent susceptibility can be expressed as $\chi_0 = \chi_p + \chi_{\mathrm{core}} + \chi_{\mathrm{vv}} + \chi_L$, where $\chi_p$ represents Pauli paramagnetism, and the remaining terms denote the diamagnetic susceptibility of the atomic core, the van Vleck paramagnetic effect, and Landau diamagnetism ($\chi_L \approx -\frac{1}{3}\chi_p$), respectively. Using Pascal's technique\cite{Rw}, $\chi_{\mathrm{core}}$ is calculated to be $1.39 \times 10^{-4} $ emu mol$^{-1}$. Since van Vleck paramagnetism is typically negligible in rare-earth ions with an even number of unpaired $4f$ electrons, $\chi_p$ is estimated to be $4.63 \times 10^{-4}$ emu mol$^{-1}$.

\begin{figure}[tbp]
\includegraphics[width=8.5cm]{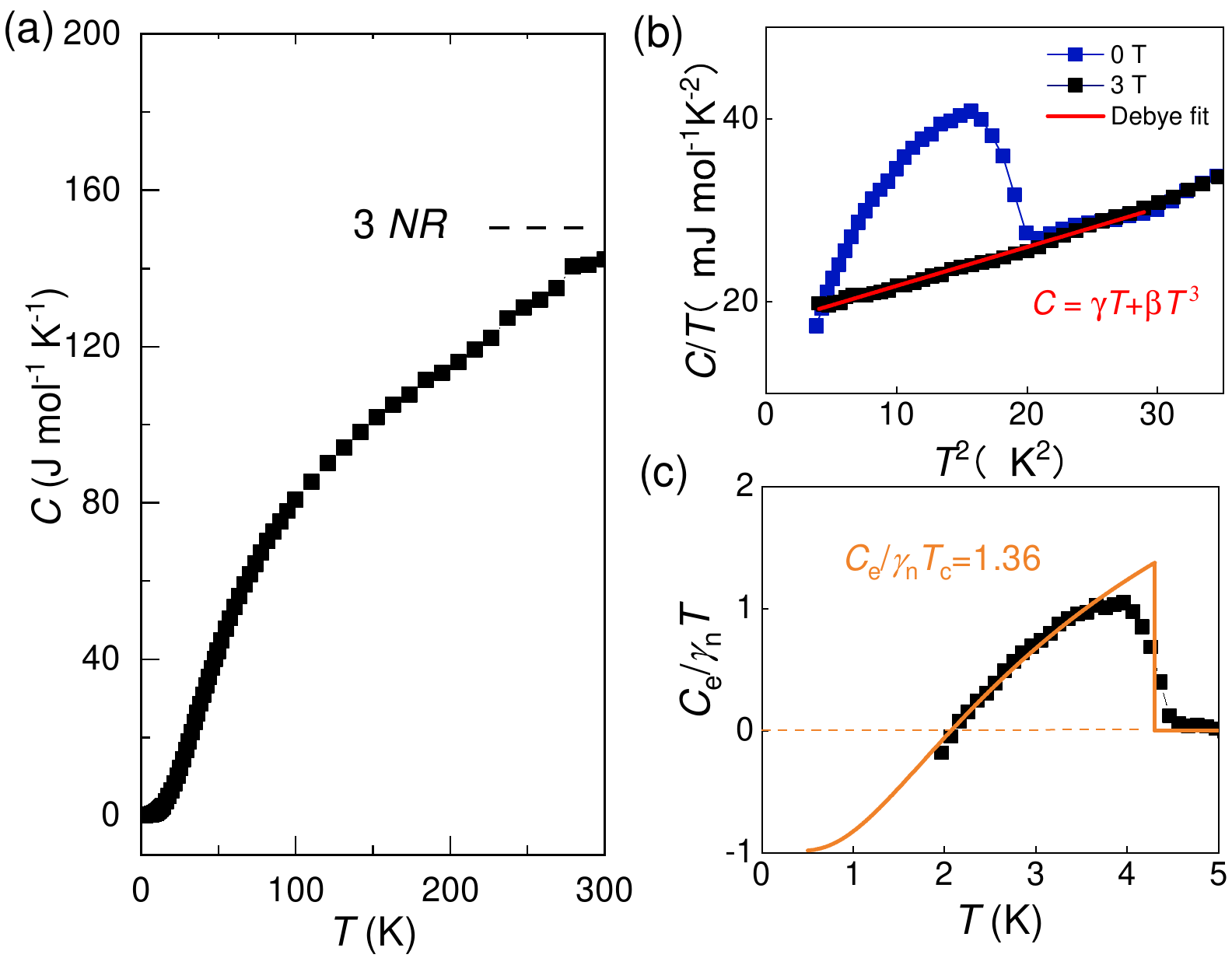}
\caption {(Color online) (a) Temperature dependence of zero-field specific heat for LuOs$_3$B$_2$ from 300 K to 1.8 K. (b) Low-temperature specific heat data at 0 T and 3 T, plotted as $C_p/T$ versus $T$, with the solid line representing a fit to the 9 T data using the Debye model. (c) Normalized electronic specific heat \(C_e / \gamma_n T_c\) after subtracting the normal state contribution. The solid red line presents a fit to the conventional BCS model.}
\label{fig4}
\end{figure}

The temperature dependence of the specific heat is presented in Figure 4(a). At high temperatures, the data agree reasonably well with the Dulong-Petit law, which predicts a value of \(3NR \approx \text{150} \text{ J mol}^{-1} \text{K}^{-1}\), where \(N\) is the number of atoms per formula unit and \(R\) is the gas constant. A distinct jump in the specific heat is observed at low temperatures, as shown in Figure 4(b), indicating the presence of bulk SC. The superconducting transition is suppressed under an applied magnetic field of 3 T. Typically, the low-temperature specific heat can be expressed as $C = \gamma T + \beta T^3$, where the first term represents the electronic contribution and the second term accounts for the lattice vibrations. By fitting the specific heat data below $5\,\mathrm{K}$ (at $B = 3\,\mathrm{T}$), we obtain the parameters $\gamma = 17.8\,\mathrm{mJ\,mol^{-1}\,K^{-2}}$ and $\beta = 0.371\,\mathrm{mJ\,mol^{-1}\,K^{-4}}$. The Debye temperature \(\Theta_D\) is estimated to be \(316.2 \, \text{K}\) by the formula $\Theta_D = \left[\frac{12\pi^4 NR}{5\beta}\right]^{1/3}$.

After subtracting the phonon contribution, the normalized electronic specific heat \(C_e / \gamma_n T_c\) is shown in Fig.~\ref{fig4}(c). The electronic specific heat jump at \(T_c\) yields \(C_e / \gamma_n T_c \sim 1.36\), further confirming bulk superconductivity in \(\text{LuOs}_3\text{B}_2\). This value is close to the theoretical BCS weak-coupling limit of \(1.43\). Additionally, the dimensionless electron-phonon coupling constant \(\lambda_{ep}\) can be estimated using the inverted McMillan equation\cite{mcmillan1968transition}:
\[
\lambda_{ep} = \frac{1.04 + \mu^* \ln(\Theta_D / 1.45 T_c)}{(1 - 0.62 \mu^*) \ln(\Theta_D / 1.45 T_c) - 1.04},
\]
where \(\mu^* = 0.13\) is the Coulomb pseudopotential for polyvalent transition metals. Evaluating this equation gives \(\lambda_{ep} = 0.61\), indicating that \(\text{LuOs}_3\text{B}_2\) exhibits moderate-coupling superconductivity.

\begin{figure}[tbp]
	\includegraphics[width=8.5cm]{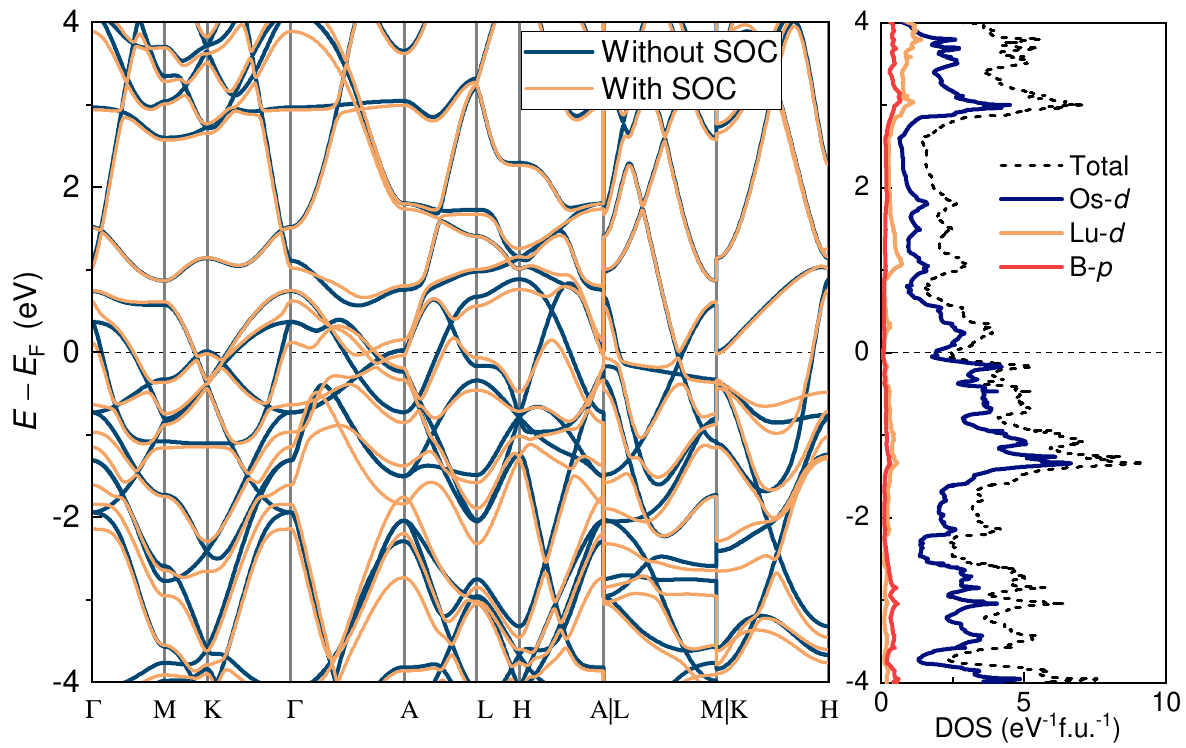}
	\caption {(Color online) First-principles electronic structure calculations of LuOs$_3$B$_2$.  }
	\label{fig5}
\end{figure}
\begin{figure}[tbp]
	\includegraphics[width=8.5cm]{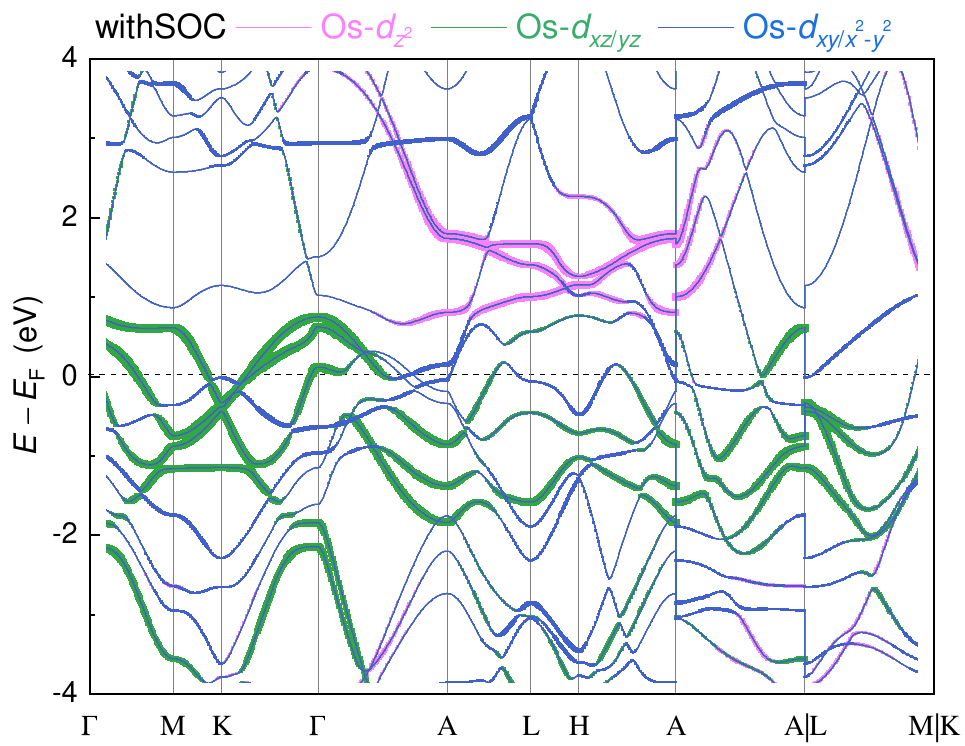}
	\caption {(Color online) The Os $d$-orbital
decomposed band structure of LuOs$_3$B$_2$.  }
	\label{fig6}
\end{figure}

The electronic structure serves as a fundamental framework for elucidating the intrinsic properties of kagome materials. First-principles calculations of LuOs$_3$B$_2$, both with and without spin-orbit coupling effects, are presented in Figure \ref{fig5}. Analysis reveals multiple bands intersecting the Fermi level, confirming the material's metallic nature. The low-energy band structure predominantly originates from Os $d$-orbital states. Consistent with minimal kagome tight-binding model predictions, LuOs$_3$B$_2$ exhibits typical kagome band features, including symmetry-protected Dirac crossings at the Brillouin zone (BZ) corner K point, van Hove singularities near the BZ boundary M point, and quasi-flat bands within the $\Gamma$-M-K plane. Figure \ref{fig6} displays the Os $d$-orbital decomposed band structure, revealing that the kagome-characteristic bands primarily originate from the Os-$d_{xz/yz}$ orbitals. This orbital dependence markedly contrasts with that observed in LaRu$_3$Si$_2$, where the corresponding bands are dominated by the Ru-$d_{z^2}$ orbital\cite{LaRuSi-1}. This distinction suggests significant differences in the electronic structure and orbital contributions between the two systems, potentially arising from the contrasting roles of the 5$d$ (Os) and 4$d$ (Ru) orbitals in shaping their respective band structures. 
Furthermore, the inclusion of SOC leads to substantial band splitting, particularly evident in the formation of distinct energy gaps at the Dirac nodes. In addition, the proximity of these  Dirac points to the Fermi level suggests potential topological surface states, though their experimental observation may be complicated by overlapping bulk states at equivalent energies. The density of states at the Fermi level, $N(E_F)$, is calculated to be approximately 2.5 states/eV/f.u.. Using this value, the band-structure Sommerfeld coefficient is estimated as $\gamma_{\text{band}} = \frac{\pi^2 k_B^2}{3} N(E_F) = 5.9$ mJ/mol/K$^2$. The experimental $\gamma$ value exceeds the band calculation by a factor of $\gamma/\gamma_{\text{band}} \approx 3$, which could suggest an electron-mass renormalization due to correlation effect. Interestingly, this  renormalization appears significantly more pronounced in Kagome superconductor CsCr$_3$Sb$_5$\cite{Cr135}.

To further quantify the electron correlation effects, the Wilson ratio\cite{wilson1975}, $R_W$, is evaluated using the expression:
\[
R_W = \frac{4\pi^2 k_B^2}{3(g\mu_B)^2} \frac{\chi_p}{\gamma},
\]
where $\chi_p$ is the Pauli susceptibility, $\gamma$ is the specific heat coefficient, $g \approx 2$ is the Landé factor, and $\mu_B$ is the Bohr magneton. The Wilson ratio is a dimensionless quantity that is about 1 for the noninteracting electron gas, while it ranges from 1 to 2 for an interacting Fermi liquid. When \( R_W > 2 \), strong electron correlation effects are expected, as typically observed in systems like Mott insulators\cite{Mott1,Mott2}. For $\text{LuOs}_3\text{B}_2$, the obtained $R_W$ value of 1.89 indicates the presence of moderate electron correlations. Note that a similar Wilson ratio has also been observed in isostructural compounds $R$Ru$_3$Si$_2$.\cite{YRSi,ThRuSi}. All relevant physical parameters of LuOs$_3$B$_2$ are summarized in Table II.

\begin{table}[h!]
\centering
\caption{Physical parameters of LuOs$_3$B$_2$ at superconducting and normal states.}
\label{tab:parameters}
\begin{tabular}{c  c c}
\hline
Parameter  & Value & Units \\ \hline
$T_{c}$  & 4.63& K \\
$\mu_{0}H_{c1}(0)$  & 13.4 & mT\\
$\mu_{0}H_{c2}(0)$ & 2.3& T  \\
$\mu_0 H_c(0)$ & 106 & mT\\
$\xi_{\rm GL}$  & 120 & Å\\
$\lambda_{\rm GL}$  & 1830& Å \\
$\kappa_{\rm GL}$  & 15.2 &\\
$\gamma$ & 17.8 & mJ$\cdot$mol$^{-1}\cdot$K$^{-2}$\\
$\Theta_{\rm D}$& 316.2 & K \\
$\Delta C_{\rm ele}/\gamma T_{c}$  & 1.36& \\
$\lambda_{\rm e-ph}$  & 0.61&\\
$R_{\rm w}$& 1.89  &\\ \hline\hline
\end{tabular}
\end{table}

Now let us discuss the superconductivity and electronic structure of the 132 family of compounds. While  preliminary studies  tend to support  that these compounds exhibit conventional phonon-mediated BCS superconductivity\cite{LaRuSiwhh,YRSi,ThRuSi,LaIrGa}, recent investigations have uncovered charge order in La(Ru$_{1-x}$Fe$_x$)$_3$Si$_2$, suggesting the potential for unconventional superconducting mechanisms\cite{LaRuFeSi}. Furthermore, enhanced electronic correlations are a prevalent feature in these systems, often attributed to the presence of flat bands near the Fermi level in Ru-based compounds\cite{LaRuSi-1,ThRuSi,YRSi}. In contrast, for LuOs$_3$B$_2$, the flat bands are situated far from the Fermi level, and the observed electronic correlations are likely driven by van Hove singularities and strong spin-orbit coupling. These characteristics stem from the perfect ideal kagome lattice and the extended nature of the Os-5$d$ orbitals. Notably, the SOC-induced gap opening at the Dirac points further enhances the potential of LuOs$_3$B$_2$ to host topologically protected superconductivity. To fully explore this possibility, the growth of high-quality single crystals is essential for advancing our understanding. Future experimental efforts combining $\mu$SR, STM, and ARPES will be crucial to elucidate the interplay between electronic correlations, topological states, and superconductivity in this system.

\section{\label{sec:level4} CONCLUSIONS}

In conclusion, our integrated experimental and theoretical study of the kagome metal LuOs$_3$B$_2$ reveals a multifaceted interplay of superconductivity, electron correlations, and topological band features. Detailed analysis of resistivity, magnetization, and specific heat data reveals a bulk type-II superconducting phase below $T_c = 4.63$ K, exhibiting a moderate electron-phonon coupling strength ($\lambda_{ep} = 0.61$) and an upper critical field $\mu_0 H_{c2}(0) = 2.3$ T. In the normal state, the susceptibility follows a Curie-Weiss behavior, while the electrical resistance demonstrates Fermi-liquid characteristics at low temperatures. On the other hand, the enhanced Wilson ratio suggests the presence of moderate electron correlation effects in the system. First-principles calculations reveal the electronic structure dominated by Os $d$-orbital derived states near $E_F$ with nontrivial topological features arising from the kagome lattice geometry. This work not only advances the fundamental understanding of 132-type kagome superconductors but also positions them as a promising class of materials for probing the interplay of quantum phenomena in correlated topological systems.

\begin{center}
\textbf{ACKNOWLEDGEMENTS}
\end{center}
This work was supported by the National Natural Science
Foundation of China (Grant no. 11104224), the Sichuan
Science and Technology Development Project (no.
2021ZYD0027), and  the Sichuan Natural Science Foundation(no. 2022NSFSC0340). We gratefully acknowledge support by the Foundation of Key Laboratory of Magnetic Suspension Technology and Maglev Vehicle, Ministry of Education.


\begin{thebibliography}{99}
\bibitem{available}
	All data are available from the authors upon reasonable request
\end{thebibliography}
\end{document}